\documentclass[aps,prb,reprint,english,a4paper,floatfix,showpacs,10pt,superscriptaddress,longbibliography]{revtex4-1}
\usepackage[T1]{fontenc}
\usepackage{ae,aecompl}
\usepackage[english]{babel}
\usepackage{mathtools}
\usepackage[normalem]{ulem}
\usepackage{url}
\usepackage{psfrag,color,pstricks,pst-grad}
\usepackage[dvips]{graphicx}
\usepackage{amsmath,amsfonts,amssymb,amsbsy}  
\usepackage{hyperref}

\newcommand{\ca}{\mathcal}

\newcommand{\beq}{\begin{equation}}
\newcommand{\eeq}{\end{equation}}

\begin{document}

\title{Vibrational mean free paths and thermal conductivity of amorphous silicon from non-equilibrium molecular dynamics simulations}
\author{K. S\"a\"askilahti}
\email{kimmo.saaskilahti@aalto.fi}
\affiliation{Engineered Nanosystems group, School of Science, Aalto University, P.O. Box 12200, 00076 Aalto, Finland}
\affiliation{Department of Mechanical Engineering, Carnegie Mellon University, 5000 Forbes Avenue, Pittsburgh 15213, United States}
\author{J. Oksanen}
\author{J. Tulkki}
\affiliation{Engineered Nanosystems group, School of Science, Aalto University, P.O. Box 12200, 00076 Aalto, Finland}
\author{A. J. H. McGaughey}
\affiliation{Department of Mechanical Engineering, Carnegie Mellon University, 5000 Forbes Avenue, Pittsburgh 15213, United States}
\author{S. Volz}
\email{sebastian.volz@centralesupelec.fr}
\affiliation{\'Ecole Centrale Paris, Grande Voie des Vignes, 92295 Ch\^atenay-Malabry, France}
\affiliation{CNRS, UPR 288 Laboratoire d'Energ\'etique Mol\'eculaire et Macroscopique, Combustion (EM2C), Grande Voie des Vignes, 92295 Ch\^atenay-Malabry, France}

\date{\today}
 
\begin{abstract}
The frequency-dependent mean free paths (MFPs) of vibrational heat carriers in amorphous silicon are predicted from the length dependence of the spectrally decomposed heat current (SDHC) obtained from non-equilibrium molecular dynamics simulations. The results suggest a (frequency)$^{-2}$ scaling of the room-temperature MFPs below 5 THz. The MFPs exhibit a local maximum at a frequency of 8 THz and fall below 1 nm at frequencies greater than 10 THz, indicating localized vibrations. The MFPs extracted from sub-10 nm system-size simulations are used to predict the length-dependence of thermal conductivity up to system sizes of 100 nm and good agreement is found with separate molecular dynamics simulations. Weighting the SDHC by the frequency-dependent quantum occupation function provides a simple and convenient method to account for quantum statistics and provides reasonable agreement with the experimentally-measured trend and magnitude.
\end{abstract}
 \maketitle

\section{Introduction} 

Compared to heat transfer by phonons in crystalline materials, heat transfer in amorphous materials is complicated by the existence of three regimes of vibrational modes.\cite{allen99} Low-frequency propagons are delocalized and have a well-defined wave vector and group velocity,\cite{feldman93} similar to phonons in crystals, while high-frequency locons are localized and contribute negligibly to thermal conduction.\cite{leitner01} Diffusons have intermediate frequencies and are delocalized, but do not have well-defined wave vectors or group velocities. The contribution of diffusons to thermal conduction can be notable, however, as they occupy the majority of the vibrational spectrum.\cite{feldman93} 

From kinetic theory,\cite{ziman} the contribution of an individual phonon or propagon mode to thermal conductivity is proportional to its mean free path (MFP). Because diffusons do not have a well-defined group velocity, it is not clear if they have a MFP or how it can be defined. Their contribution to thermal conductivity can be predicted using their diffusivity, which is well-defined and can be calculated from Allen-Feldman theory.\cite{allen93,feldman93} Nevertheless, it would be insightful to identify a frequency-dependent length scale for propagons and diffusons describing the decay of the heat flux at each vibrational frequency. Such a definition would lift the (fundamentally) arbitrary distinction between propagons and diffusons and enable a unified description of heat transfer at all vibrational frequencies. 

In this paper, we apply the spectrally-decomposed MFP method\cite{saaskilahti15} to probe the non-equilibrium MFPs of vibrational heat carriers in amorphous silicon (a-Si). This method is based on calculating the spectrally-decomposed heat current (SDHC)\cite{saaskilahti14b}  in systems of different lengths using non-equilibrium molecular dynamics (NEMD) simulations. The MFPs are determined from the variation of the SDHC as a function of system length at each vibrational frequency.\cite{saaskilahti15b} 

We previously used the spectrally-decomposed MFP method to calculate the non-equilibrium MFPs in low-dimensional systems such as carbon nanotubes\cite{saaskilahti15} and anharmonic chains.\cite{saaskilahti15b} We demonstrated that the non-equilibrium MFPs transparently describe the ballistic-to-diffusive transition in the length-dependence of thermal conductivity. Compared to previous calculations for a-Si, the spectrally-decomposed MFP method has several advantages. Unlike in modal life-time calculations,\cite{he11} we do not need to estimate the group velocities of individual modes to calculate their MFPs. We also do not need to distinguish between propagons and diffusons\cite{larkin14} nor resort to the harmonic approximation.\cite{feldman93} In contrast to recent calculations studying the spectral conductivity of a-Si in fixed-size systems, \cite{lv15,zhou15} we focus on the MFPs of heat carriers and the system-size dependence of thermal conductivity. 

The rest of the paper is organized as follows. The calculation methods are presented in Sec. \ref{sec:setup} and the numerical results are discussed in Sec. \ref{sec:results}. We also introduce a simple method for the quantum correction of thermal conductivity from classical MD simulations, based on weighting the SDHC by the quantum occupation function. Because this quantum correction method operates at the frequency level, it is more reasonable than quantum-correction methods that operate at the system level (see Ref. \citenum{turney09a} and references therein) and allows us to compare our predictions to experimental measurements. 

\section{Simulation setup and methods} \label{sec:setup}

\begin{figure}[t]
 \begin{center}
  \includegraphics[width=8.6cm]{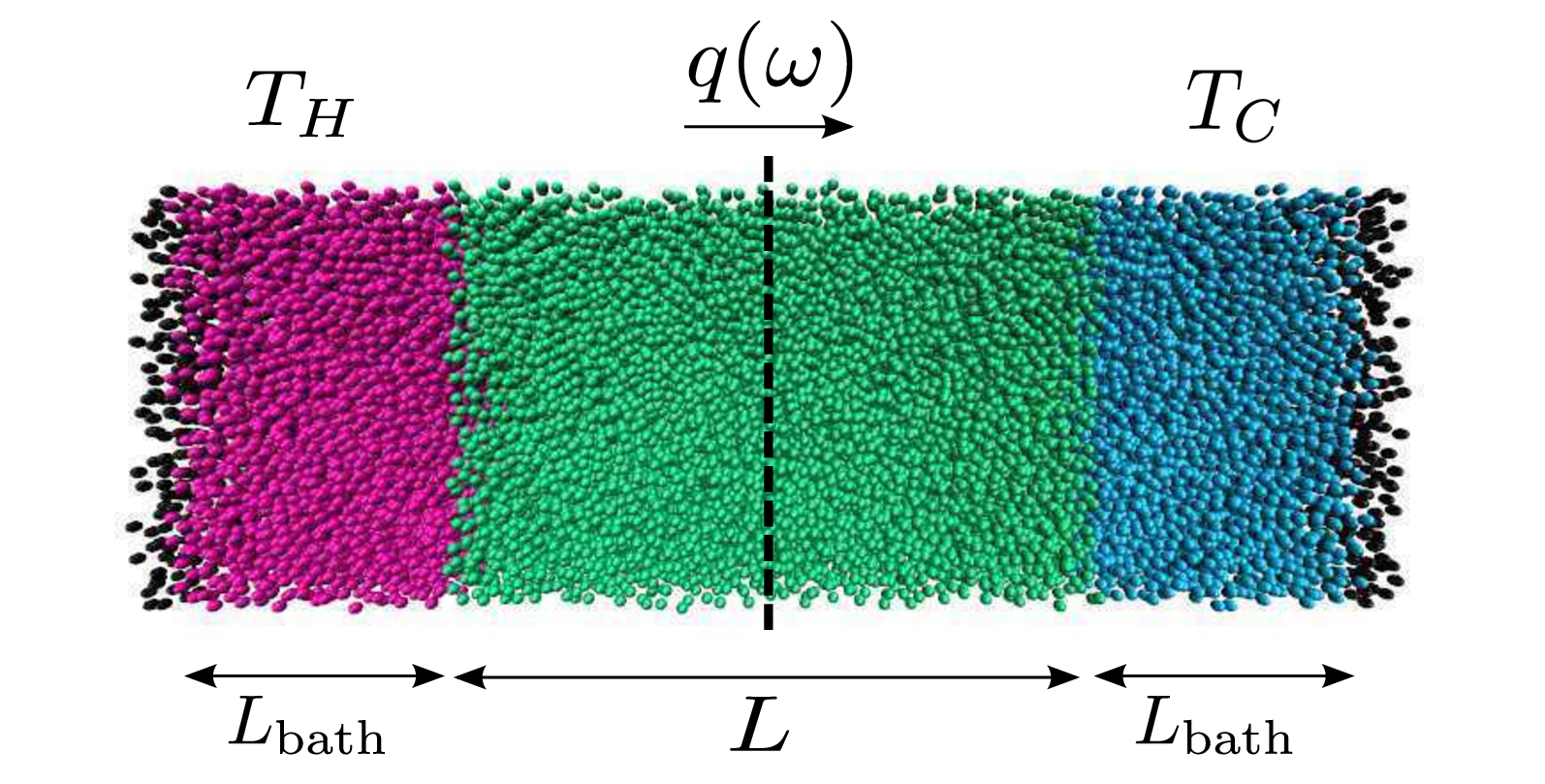}
 \caption{Schematic illustration of the a-Si system for $L=10$ nm. The spectral heat flux $q(\omega)$ is calculated at the cross-section in the middle of the structure (dashed line). The length $L$ between the Langevin heat baths is varied to extract the vibrational mode MFPs  based on the decrease of $q(\omega)$ as a function of $L$.}
 \label{fig:geom}
 \end{center}
\end{figure}

All simulations are carried out using the LAMMPS package\cite{plimpton95} with a time step  of 2.5 fs. The Si-Si interactions are modeled by the Stillinger-Weber potential\cite{stillinger85} with the parameters of Ref. \citenum{vink01}. The NEMD simulation geometry is shown in Fig. \ref{fig:geom}. The atomic coordinates for a-Si are generated by following the melt-quench procedure of Ref. \citenum{france-lanord14} and the final density is 2,291 kg/m$^3$.  After the equilibration of the quenched system, atoms located within a distance $L_{\textrm{bath}}=5$ nm from the left and right edges of the structure are coupled to Langevin heat baths at temperatures $T_H=T+\Delta T/2$ and $T_C=T-\Delta T/2$ with bath relaxation times of 1 ps. To prevent sublimation, atoms at the far left and right edges are fixed to their equilibrium positions. Periodic boundary conditions are applied at the boundaries transverse to the current flow. The width of the system cross-section is $7$ nm. System lengths $L$ (i.e., the region between the baths) between 1 and 10 nm at intervals of 1 nm are considered.

The SDHC is calculated through the plane of decomposition located halfway between the hot and cold baths (dashed line in Fig. \ref{fig:geom}).\cite{saaskilahti15,saaskilahti15b} The SDHC $q_{i \to j}(\omega)$ between particles $i$ and $j$ located on opposite sides of this plane is given by the pair-wise SDHC equation \cite{saaskilahti14b}
 \begin{equation}
 q_{i \to j}(\omega) = -\frac{2}{t_{\textrm{simu}} \omega} \sum_{\alpha,\beta\in\{x,y,z\}} \textrm{Im}\left.\left\langle \hat{v}_i^{\alpha}(\omega)^* K_{ij}^{\alpha\beta} \hat{v}_j^{\beta}(\omega) \right\rangle \right. \label{eq:qomega_expr},
\end{equation}
where $t_{\textrm{simu}}$ is the simulation time, $\omega$ is the angular frequency, and the interatomic force constant $K_{ij}^{\alpha\beta}$ is defined as 
\begin{equation}
 K_{ij}^{\alpha\beta} = - \left. \frac{\partial ^2 \mathcal{V}}{\partial u_i^{\alpha} u_j^{\beta}} \right|_{\mathbf{u}=\mathbf{0}}. \label{eq:K}
\end{equation}
The velocities $\hat v_i^{\alpha}(\omega)$ and $\hat v_j^{\beta}(\omega)$ are the discrete Fourier transforms of the atomic velocities $v_i^{\alpha}(t)=\dot{u}_i^{\alpha}(t)$ and $v_j^{\beta}(t)=\dot{u}_j^{\beta}(t)$ (the exact definitions are in Ref. \citenum{saaskilahti15}), where $u_i^{\alpha}$ and $u_j^{\beta}$ are the displacements of atoms $i$ and $j$ from their equilibrium positions in directions $\alpha$,$\beta \in \{x,y,z\}$. In Eq.~\eqref{eq:K}, $\ca{V}$ is the interatomic potential energy function. The spectral flux through the plane of decomposition is obtained from Eq.~\eqref{eq:qomega_expr} by summing over all pairs of atoms (one of the left side, denoted by $\tilde{L}$, and one on the right side, denoted by $\tilde{R}$) within the potential cut-off distance of each other and dividing by the interface area $A$: 
\begin{equation}
 q(\omega) = \frac{1}{A} \sum_{i\in  \tilde L} \sum_{j\in \tilde R} q_{i \to j}(\omega). \label{eq:qomega_sum}
\end{equation}
While Eq.~\eqref{eq:qomega_expr} is the first-order approximation to the inter-particle SDHC,\cite{saaskilahti14b} we have confirmed that the contribution of higher-order terms is negligible for a-Si by comparing the integral of Eq. \eqref{eq:qomega_sum}, which we denote as $Q$, to the total flux determined from the work done by the heat baths. The two results agree within 4\%. We attribute this good agreement to the stiffness of the interatomic bonds in a-Si, which ensures that the first-order term in the current (proportional to the harmonic force constants $K_{ij}^{\alpha \beta}$) dominates the higher-order terms that are related to anharmonic force constants. This restriction to the first-order current term at the plane of decomposition does not, however, mean that anharmonic scattering in the bulk is neglected, because all anharmonic effects are included in the NEMD simulations.\cite{saaskilahti14b,saaskilahti15}

Frequency-dependent MFPs $\Lambda(\omega)$ are calculated by determining $q(\omega,L)$ for different system lengths $L$ and fitting the length-dependent $q(\omega,L)$ to the equation \cite{saaskilahti15,saaskilahti15b}
\begin{equation}
 q(\omega,L) = \frac{q^0(\omega)}{1+L/[2\Lambda(\omega)]}, \label{eq:mfp}
\end{equation}
where $q^0(\omega)$ is the spectral flux when the baths are in contact ($L\to 0^+$). Both $q^0(\omega)$ and $\Lambda(\omega)$ are determined from the fitting procedure. While $q^0(\omega)$ depends on the details of the heat baths, the MFPs extracted from the length-dependence are not expected to depend on the bath details.\cite{saaskilahti15b} The frequency-dependent MFPs determined from Eq. \eqref{eq:mfp} are mode-averaged and projected along the direction of heat transfer.\cite{saaskilahti15} The MFPs are independent of the system length and therefore correspond to the bulk values.\cite{saaskilahti15} We note that the MFPs determined from Eq. \eqref{eq:mfp} correspond to the decay length of the heat flux. This definition does not necessarily coincide with the conventional definition of the MFP as the decay length of a wave packet.\cite{ziman}  This distinction is important for diffusons, which do not have a well-defined wave-vector so that the traditional definition cannot be applied.

Once the spectral MFPs are determined, the thermal conductivity $\kappa$ for length $L$ can be determined from\cite{saaskilahti15}
\begin{equation}
 \kappa = \frac{QL}{\Delta T}= \frac{L}{\Delta T} \int_0^{\infty} \frac{d\omega}{2\pi} \frac{q^0(\omega)}{1+L/[2\Lambda(\omega)]}. \label{eq:kappa_est}
\end{equation}
The length-dependence of the thermal conductivity can be intuitively understood by writing Eq. \eqref{eq:kappa_est} in the equivalent form
\begin{equation}
 \kappa = \frac{2}{\Delta T} \int_0^{\infty} \frac{d\omega}{2\pi} q^0(\omega) \underbrace{\frac{1}{(L/2)^{-1}+\Lambda(\omega)^{-1}}}_{\Lambda_{\textrm{eff}}(\omega)^{-1}},
\end{equation}
where the ``effective'' MFP $\Lambda_{\textrm{eff}}(\omega)$ has been introduced, which is similar to the well-known Matthiessen rule.\cite{chen} The effective MFP accounts for boundary scattering through the additional $L/2$ term and is limited to below this value.

Finally, Eq.~\eqref{eq:kappa_est} allows for a simple quantum correction to the thermal conductivity prediction, because the contributions of different frequencies can be weighted by the vibrational mode energy and occupation,
as in the Landauer-B\"uttiker formalism \cite{rego98,angelescu98,saaskilahti13}. We define the quantum corrected thermal conductivity as
\begin{equation}
 \kappa = \frac{L}{\Delta T} \int_0^{\infty} \frac{d\omega}{2\pi} \frac{q^0(\omega)}{1+L/[2\Lambda(\omega)]} \frac{\hbar \omega}{k_\mathrm{B}} \frac{\partial f_{\mathrm {BE}}(\omega,T)}{\partial T}, \label{eq:kappa_quantum}
\end{equation}
where $f_\mathrm{BE}(\omega,T)=\left[ \exp(\hbar \omega/k_\mathrm{B} T)-1\right]^{-1}$ is the Bose-Einstein distribution function, $k_\mathrm{B}$ is the Boltzmann constant, and $\hbar$ is the Planck constant divided by $2\pi$. By defining the dimensionless, length-dependent bath-to-bath transmission function as
\begin{equation}
 \ca{T}(\omega,L) = \frac{q^0(\omega)A}{k_\mathrm{B} \Delta T} \frac{1}{1+L/[2\Lambda(\omega)]},
\end{equation}
Eq. \eqref{eq:kappa_quantum} can be written in the familiar Landauer-B\"uttiker form as \cite{rego98}
\begin{equation}
 \kappa = \frac{L}{A} \int_0^{\infty} \frac{d\omega}{2\pi} \hbar \omega \ca{T}(\omega,L)  \frac{\partial f_B(\omega,T)}{\partial T}. \label{eq:kappa_quantum2}
\end{equation}
The proposed quantum correction accounts for the quantum specific heat of the modes at each frequency, but does not account for quantum effects in the dynamics. The method is thus similar to the one recently introduced by Lv and Henry,\cite{lv15} who weight the modal contributions to the equilibrium Green-Kubo thermal conductivity by the quantum population function.

All our NEMD simulations are performed at a mean temperature of $T=300$ K with temperature bias $\Delta T=100$ K. Choosing a relatively large temperature bias allows for very good signal-to-noise ratio in the spectral heat flux, suppressing the statistical noise. We checked that halving the bias to $\Delta T=50$ K does not change the spectral MFPs. In addition, we checked that the heat flux is not sensitive to the exact arrangement of atoms arising from the melt-and-quench procedure, which we attribute to the large cross-section of the system giving rise to spatial averaging in the spectral currents. Therefore, we performed a single melt-quench for each system length.

\section{Results}\label{sec:results}

\begin{figure}[t]
 \begin{center}
  \includegraphics[width=8.6cm]{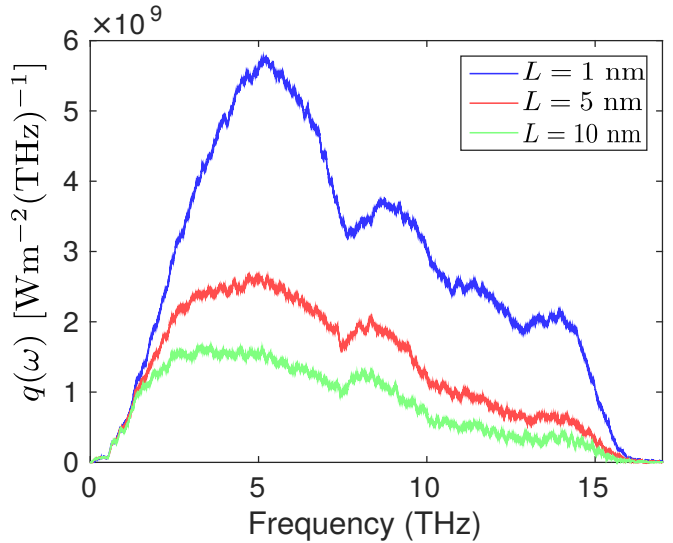}
 \caption{Spectral heat flux $q(\omega)$ for different system lengths $L$. Increasing the system length reduces the heat flux, especially at high frequencies, where the MFPs are shorter. At frequencies below 2 THz, the heat current is independent of system length, suggesting ballistic transport.}
 \label{fig:flux}
 \end{center}
\end{figure}

The spectral heat flux $q(\omega)$ for selected system lengths $L$ as a function of frequency $f=\omega/(2\pi)$ is plotted in Fig. \ref{fig:flux}. The spectral distribution of the heat flux for $L=20$ nm was recently analyzed in detail by Zhou and Hu,\cite{zhou15} so we focus here on its length-dependence. As expected, increasing the system length reduces the heat current throughout the whole frequency range because of increased phonon-phonon scattering. The reduction is strongest at high frequencies, where the MFPs are shorter compared to low frequencies. At frequencies less than 2 THz, the spectral current is nearly independent of system length. Such nearly ballistic conduction suggest that the low frequency MFPs are notably longer than the system sizes considered here. 

Equation \eqref{eq:mfp} suggests that the inverse of the spectral flux will be linearly proportional to the system length, with the slope given by $[2\Lambda(\omega)]^{-1}$. To determine the MFPs, we calculated the spectral flux for system sizes $L \in \{1,2,\dots,10\}$ nm, plotted $q(\omega)^{-1}$ versus $L$ and fitted a linear function at each frequency using least squares fitting.\cite{saaskilahti15,saaskilahti15b} A linear function accurately reproduces the length-dependence of $q(\omega)^{-1}$ (not shown), as previously also observed for other systems.\cite{saaskilahti15,saaskilahti15b}

\begin{figure}[t]
 \begin{center}
  \includegraphics[width=8.6cm]{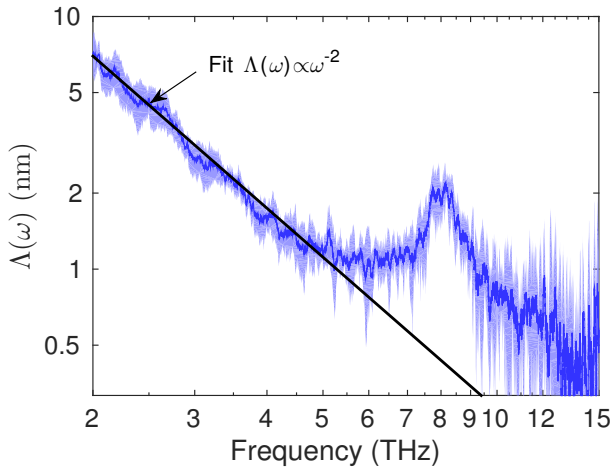}
 \caption{Log-log plot of the spectral MFPs determined by fitting to Eq. \eqref{eq:mfp}. The shaded regions correspond to the 95 \% confidence interval.}
 \label{fig:mfps}
 \end{center}
\end{figure}

Because of the high computational cost associated with calculating the spectral heat fluxes for large systems, we limited our study to systems at most 10 nm long. This restriction precludes extracting MFPs longer than 10 nm accurately, limiting the current analysis to frequencies greater than 2 THz. The MFPs $\Lambda(\omega)$ extracted from the linear fitting procedure are shown in a log-log plot in Fig.~\ref{fig:mfps}. At frequencies below 5 THz, the MFPs obey a power-law scaling $\Lambda(\omega) \propto \omega^{-2}$. This scaling agrees with modal life-time calculations on a-Si.\cite{larkin14} At frequencies greater than 5 THz, the power-law scaling breaks down and the MFPs increase with increasing frequency, giving rise to a local maximum around 8 THz. A similar maximum for a-Si has been observed in effective MFPs \cite{he11} and in lifetimes.\cite{larkin14} At higher frequencies, the MFPs decrease again and fall below 1 nm, which is on the order of the silicon-silicon bond length. At such high frequencies, the uncertainty is large because of the sensitivity of the spectral flux to the system size.

Larkin and McGaughey reported a propagon-diffuson transition frequency of 1.8 THz,\cite{larkin14} such that the frequency range considered in Fig. \ref{fig:mfps} mostly corresponds to diffuson-like modes. In the analysis below, we assume that the scaling $\Lambda(\omega) \propto \omega^{-2}$ (solid line in Fig. \ref{fig:mfps}) remains valid at frequencies below 2 THz. With such scaling, the MFPs exceed 100 nm below 530 GHz and 1 $\mu$m below 170 GHz. While the $\omega^{-2}$ scaling may break down in real situations because of defects, boundary scattering, or even the onset of  a Rayleigh-like $\omega^{-4}$ scaling at very low frequencies, \cite{feldman93} we assume it to hold for simplicity.

\begin{figure}[t]
 \begin{center}
  \includegraphics[width=8.6cm]{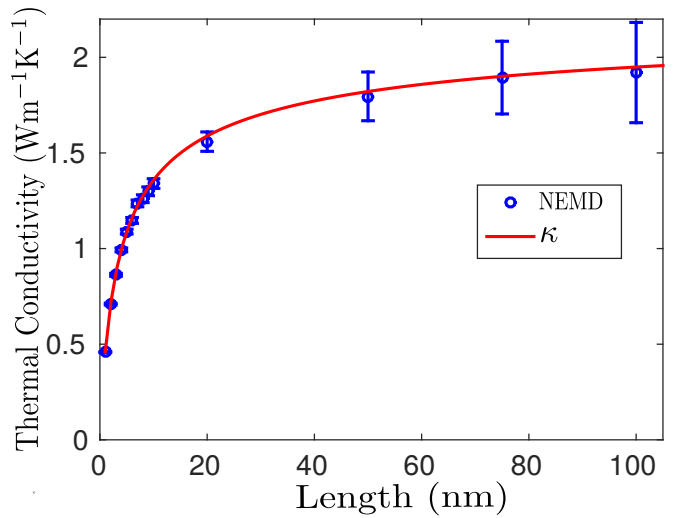}
 \caption{Thermal conductivity versus system length. The thermal conductivities calculated from direct NEMD simulation are marked by circles and the estimated thermal conductivity from Eq.~\eqref{eq:kappa_est} using classical statistics is indicated by the solid line. The error bars in the NEMD thermal conductivities correspond to the 95\% confidence interval.}
 \label{fig:kappa}
 \end{center}
\end{figure}


We now investigate the length-dependence of the thermal conductivity using Eq.~\eqref{eq:kappa_est}. The calculated thermal conductivity (continuous line) is compared with that determined directly from NEMD simulations (data points) for lengths up to $100$ nm in Fig.~\ref{fig:kappa}. In the evaluation of the integral in Eq.~\eqref{eq:kappa_est}, the MFPs have been assumed to scale as $\Lambda(\omega)\propto \omega^{-2}$ at frequencies below 2 THz. 
These calculations were carried out without the quantum correction as the NEMD simulations are classical. Equation \eqref{eq:kappa_est} combined with the MFP data of Fig. \ref{fig:mfps} reproduces the length-dependence of thermal conductivity up to lengths $L=100$ nm to within 2\%. This close agreement (i) supports the assumption of $\Lambda(\omega)\propto \omega^{-2}$ scaling at low frequencies, (ii) shows that the MFP data in Fig. \ref{fig:mfps}, which were determined from simulations of systems shorter than $10$ nm, can be reliably used to estimate the relative contributions of different vibrational frequencies to thermal transport in much larger systems, and (iii)  provides support for the accuracy of Eq. \eqref{eq:kappa_est} in describing the length-dependence.

Because the Debye temperature of a-Si is $530$ K \cite{mertig84}, which is well above room temperature, we need to apply the quantum correction to compare the predicted temperature-dependence of thermal conductivity to experimental data. To do so, we use Eq.~\eqref{eq:kappa_quantum} and evaluate the integral as a function of temperature using the MFPs from Fig. \ref{fig:mfps}, again assuming the scaling $\Lambda(\omega)\propto \omega^{-2}$ at frequencies below 2 THz. A full quantum-corrected analysis would require determining the MFPs at each temperature. For simplicity and based on the recent results of Lv and Henry \cite{lv15}, we assume that the MFPs calculated at a temperature of 300 K remain valid at other temperatures. 

 \begin{figure}[t]
  \begin{center}
   \includegraphics[width=8.6cm]{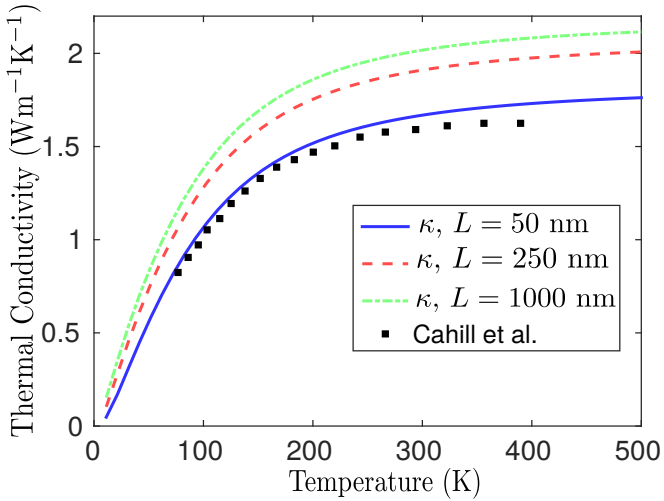}
  \caption{Quantum-corrected thermal conductivity versus temperature for system lengths of 50, 250, and 1000 nm.  The MFPs are assumed to scale as $\Lambda(\omega)\propto \omega^{-2}$ below frequencies of 2 THz and to be independent of temperature. Also plotted is the thermal conductivity of a 520 nm thick hydrogenated a-Si thin film measured by Cahill \textit{et al.} \cite{cahill94}.}
  \label{fig:kappa_qm}
  \end{center}
 \end{figure}

The quantum-corrected thermal conductivity [Eq. \eqref{eq:kappa_quantum}] is plotted as a function of temperature for system lengths of 50, 250, and 1000 nm in Fig.~\ref{fig:kappa_qm}. As noted above, assuming finite $L$ in Eq. \eqref{eq:kappa_quantum} limits the MFPs to $L/2$. Experimental data from Cahill {\em et al.} for a 520 nm thick film of hydrogenated a-Si with one atomic percent hydrogen content are also plotted.\cite{cahill94} Because available thermal conductivity measurements for a-Si contain significant scatter,\cite{larkin14} we use the data of Cahill \textit{et al.} to check the trend of our predictions, but do not expect agreement. Differences may also exist due to the use of the Stillinger-Weber potential and the classical nature of the NEMD simulations. The increase of thermal conductivity with increasing temperature is well described by the quantum-corrected thermal conductivity. At temperatures higher than 300 K, the experimentally measured thermal conductivity increases slightly slower as a function of temperature than our prediction, but this disagreement may be related to our approximation that the MFPs are independent of temperature. At such high temperatures, anharmonic scattering will reduce MFPs and therefore decrease the thermal conductivity. Without the quantum-correction, the predicted thermal conductivity would depend only very weakly on temperature (due to the weak temperature-dependence of the MFPs), precluding reasonable agreement with the trend of the experimental data.  We caution that this quantum correction has only been examined for a-Si here and that its application to other systems warrants further investigation.


\section{Conclusion} 
\label{sec:conclusion}
We investigated vibrational heat transfer in a-Si by determining the SDHC and MFPs from NEMD simulations. The calculated MFPs directly reflect the decay of the heat flux at each vibrational frequency and do not rely on the existence of a well-defined modal wave vector, thereby avoiding the separate treatment of diffusons and propagons. As shown in Fig. \ref{fig:mfps}, the MFPs exhibit $\omega^{-2}$ scaling at frequencies above 2 THz and below 5 THz. At frequencies higher than 10 THz, the MFPs fall below 1 nm, corresponding to strongly localized vibrations. The length-independent MFPs can be used to accurately predict the thermal conductivity in systems as long as 100 nm (Fig. \ref{fig:kappa}). Weighting the SDHC by the frequency-dependent quantum occupation function provides a simple method for a quantum-correction of thermal conductivity and is able to reproduce the experimentally measured temperature-dependence of thermal conductivity, as shown in Fig. \ref{fig:kappa_qm}.

In the future, it would be useful to calculate the SDHC for systems longer than those considered here, enabling direct extraction of MFPs at frequencies below 2 THz. Such an analysis could inform the ongoing discussion\cite{larkin14} of the low-frequency scaling of MFPs in a-Si. It will also be important to compare the non-equilibrium MFPs to those calculated from equilibrium molecular dynamics simulations.

\section{Acknowledgements}
This work was initiated during the research visit of K.S. to Carnegie Mellon University. The computational resources were provided by the Finnish IT Center for Science and Aalto Science-IT project. The work was partially funded by the Aalto Energy Efficiency Research Programme (AEF) and the Academy of Finland.


\begin{thebibliography}{23}%
\makeatletter
\providecommand \@ifxundefined [1]{%
 \@ifx{#1\undefined}
}%
\providecommand \@ifnum [1]{%
 \ifnum #1\expandafter \@firstoftwo
 \else \expandafter \@secondoftwo
 \fi
}%
\providecommand \@ifx [1]{%
 \ifx #1\expandafter \@firstoftwo
 \else \expandafter \@secondoftwo
 \fi
}%
\providecommand \natexlab [1]{#1}%
\providecommand \enquote  [1]{``#1''}%
\providecommand \bibnamefont  [1]{#1}%
\providecommand \bibfnamefont [1]{#1}%
\providecommand \citenamefont [1]{#1}%
\providecommand \href@noop [0]{\@secondoftwo}%
\providecommand \href [0]{\begingroup \@sanitize@url \@href}%
\providecommand \@href[1]{\@@startlink{#1}\@@href}%
\providecommand \@@href[1]{\endgroup#1\@@endlink}%
\providecommand \@sanitize@url [0]{\catcode `\\12\catcode `\$12\catcode
  `\&12\catcode `\#12\catcode `\^12\catcode `\_12\catcode `\%12\relax}%
\providecommand \@@startlink[1]{}%
\providecommand \@@endlink[0]{}%
\providecommand \url  [0]{\begingroup\@sanitize@url \@url }%
\providecommand \@url [1]{\endgroup\@href {#1}{\urlprefix }}%
\providecommand \urlprefix  [0]{URL }%
\providecommand \Eprint [0]{\href }%
\providecommand \doibase [0]{http://dx.doi.org/}%
\providecommand \selectlanguage [0]{\@gobble}%
\providecommand \bibinfo  [0]{\@secondoftwo}%
\providecommand \bibfield  [0]{\@secondoftwo}%
\providecommand \translation [1]{[#1]}%
\providecommand \BibitemOpen [0]{}%
\providecommand \bibitemStop [0]{}%
\providecommand \bibitemNoStop [0]{.\EOS\space}%
\providecommand \EOS [0]{\spacefactor3000\relax}%
\providecommand \BibitemShut  [1]{\csname bibitem#1\endcsname}%
\let\auto@bib@innerbib\@empty
\bibitem [{\citenamefont {Allen}\ \emph {et~al.}(1999)\citenamefont {Allen},
  \citenamefont {Feldman}, \citenamefont {Fabian},\ and\ \citenamefont
  {Wooten}}]{allen99}%
  \BibitemOpen
  \bibfield  {author} {\bibinfo {author} {\bibfnamefont {Philip~B.}\
  \bibnamefont {Allen}}, \bibinfo {author} {\bibfnamefont {Joseph~L.}\
  \bibnamefont {Feldman}}, \bibinfo {author} {\bibfnamefont {Jaroslav}\
  \bibnamefont {Fabian}}, \ and\ \bibinfo {author} {\bibfnamefont {Frederick}\
  \bibnamefont {Wooten}},\ }\bibfield  {title} {\enquote {\bibinfo {title}
  {Diffusons, locons and propagons: {C}haracter of atomic vibrations in
  amorphous {Si}},}\ }\href {\doibase 10.1080/13642819908223054} {\bibfield
  {journal} {\bibinfo  {journal} {Phil. Mag. Part B}\ }\textbf {\bibinfo
  {volume} {79}},\ \bibinfo {pages} {1715--1731} (\bibinfo {year}
  {1999})}\BibitemShut {NoStop}%
\bibitem [{\citenamefont {Feldman}\ \emph {et~al.}(1993)\citenamefont
  {Feldman}, \citenamefont {Kluge}, \citenamefont {Allen},\ and\ \citenamefont
  {Wooten}}]{feldman93}%
  \BibitemOpen
  \bibfield  {author} {\bibinfo {author} {\bibfnamefont {Joseph~L.}\
  \bibnamefont {Feldman}}, \bibinfo {author} {\bibfnamefont {Mark~D.}\
  \bibnamefont {Kluge}}, \bibinfo {author} {\bibfnamefont {Philip~B.}\
  \bibnamefont {Allen}}, \ and\ \bibinfo {author} {\bibfnamefont {Frederick}\
  \bibnamefont {Wooten}},\ }\bibfield  {title} {\enquote {\bibinfo {title}
  {Thermal conductivity and localization in glasses: Numerical study of a model
  of amorphous silicon},}\ }\href {\doibase 10.1103/PhysRevB.48.12589}
  {\bibfield  {journal} {\bibinfo  {journal} {Phys. Rev. B}\ }\textbf {\bibinfo
  {volume} {48}},\ \bibinfo {pages} {12589--12602} (\bibinfo {year}
  {1993})}\BibitemShut {NoStop}%
\bibitem [{\citenamefont {Leitner}(2001)}]{leitner01}%
  \BibitemOpen
  \bibfield  {author} {\bibinfo {author} {\bibfnamefont {David~M.}\
  \bibnamefont {Leitner}},\ }\bibfield  {title} {\enquote {\bibinfo {title}
  {Vibrational energy transfer and heat conduction in a one-dimensional
  glass},}\ }\href {\doibase 10.1103/PhysRevB.64.094201} {\bibfield  {journal}
  {\bibinfo  {journal} {Phys. Rev. B}\ }\textbf {\bibinfo {volume} {64}},\
  \bibinfo {pages} {094201} (\bibinfo {year} {2001})}\BibitemShut {NoStop}%
\bibitem [{\citenamefont {Ziman}(2001)}]{ziman}%
  \BibitemOpen
  \bibfield  {author} {\bibinfo {author} {\bibfnamefont {J.M.}\ \bibnamefont
  {Ziman}},\ }\href@noop {} {\emph {\bibinfo {title} {Electrons and Phonons:
  The Theory of Transport Phenomena in Solids}}}\ (\bibinfo  {publisher}
  {Oxford University Press, USA},\ \bibinfo {year} {2001})\BibitemShut
  {NoStop}%
\bibitem [{\citenamefont {Allen}\ and\ \citenamefont
  {Feldman}(1993)}]{allen93}%
  \BibitemOpen
  \bibfield  {author} {\bibinfo {author} {\bibfnamefont {Philip~B.}\
  \bibnamefont {Allen}}\ and\ \bibinfo {author} {\bibfnamefont {Joseph~L.}\
  \bibnamefont {Feldman}},\ }\bibfield  {title} {\enquote {\bibinfo {title}
  {Thermal conductivity of disordered harmonic solids},}\ }\href {\doibase
  10.1103/PhysRevB.48.12581} {\bibfield  {journal} {\bibinfo  {journal} {Phys.
  Rev. B}\ }\textbf {\bibinfo {volume} {48}},\ \bibinfo {pages} {12581--12588}
  (\bibinfo {year} {1993})}\BibitemShut {NoStop}%
\bibitem [{\citenamefont {S\"a\"askilahti}\ \emph
  {et~al.}(2015{\natexlab{a}})\citenamefont {S\"a\"askilahti}, \citenamefont
  {Oksanen}, \citenamefont {Volz},\ and\ \citenamefont
  {Tulkki}}]{saaskilahti15}%
  \BibitemOpen
  \bibfield  {author} {\bibinfo {author} {\bibfnamefont {K.}~\bibnamefont
  {S\"a\"askilahti}}, \bibinfo {author} {\bibfnamefont {J.}~\bibnamefont
  {Oksanen}}, \bibinfo {author} {\bibfnamefont {S.}~\bibnamefont {Volz}}, \
  and\ \bibinfo {author} {\bibfnamefont {J.}~\bibnamefont {Tulkki}},\
  }\bibfield  {title} {\enquote {\bibinfo {title} {Frequency-dependent phonon
  mean free path in carbon nanotubes from nonequilibrium molecular dynamics},}\
  }\href {\doibase 10.1103/PhysRevB.91.115426} {\bibfield  {journal} {\bibinfo
  {journal} {Phys. Rev. B}\ }\textbf {\bibinfo {volume} {91}},\ \bibinfo
  {pages} {115426} (\bibinfo {year} {2015}{\natexlab{a}})}\BibitemShut
  {NoStop}%
\bibitem [{\citenamefont {S\"a\"askilahti}\ \emph {et~al.}(2014)\citenamefont
  {S\"a\"askilahti}, \citenamefont {Oksanen}, \citenamefont {Tulkki},\ and\
  \citenamefont {Volz}}]{saaskilahti14b}%
  \BibitemOpen
  \bibfield  {author} {\bibinfo {author} {\bibfnamefont {K.}~\bibnamefont
  {S\"a\"askilahti}}, \bibinfo {author} {\bibfnamefont {J.}~\bibnamefont
  {Oksanen}}, \bibinfo {author} {\bibfnamefont {J.}~\bibnamefont {Tulkki}}, \
  and\ \bibinfo {author} {\bibfnamefont {S.}~\bibnamefont {Volz}},\ }\bibfield
  {title} {\enquote {\bibinfo {title} {Role of anharmonic phonon scattering in
  the spectrally decomposed thermal conductance at planar interfaces},}\ }\href
  {\doibase 10.1103/PhysRevB.90.134312} {\bibfield  {journal} {\bibinfo
  {journal} {Phys. Rev. B}\ }\textbf {\bibinfo {volume} {90}},\ \bibinfo
  {pages} {134312} (\bibinfo {year} {2014})}\BibitemShut {NoStop}%
\bibitem [{\citenamefont {S\"a\"askilahti}\ \emph
  {et~al.}(2015{\natexlab{b}})\citenamefont {S\"a\"askilahti}, \citenamefont
  {Oksanen}, \citenamefont {Volz},\ and\ \citenamefont
  {Tulkki}}]{saaskilahti15b}%
  \BibitemOpen
  \bibfield  {author} {\bibinfo {author} {\bibfnamefont {K.}~\bibnamefont
  {S\"a\"askilahti}}, \bibinfo {author} {\bibfnamefont {J.}~\bibnamefont
  {Oksanen}}, \bibinfo {author} {\bibfnamefont {S.}~\bibnamefont {Volz}}, \
  and\ \bibinfo {author} {\bibfnamefont {J.}~\bibnamefont {Tulkki}},\
  }\bibfield  {title} {\enquote {\bibinfo {title} {Nonequilibrium phonon mean
  free paths in anharmonic chains},}\ }\href {\doibase
  10.1103/PhysRevB.92.245411} {\bibfield  {journal} {\bibinfo  {journal} {Phys.
  Rev. B}\ }\textbf {\bibinfo {volume} {92}},\ \bibinfo {pages} {245411}
  (\bibinfo {year} {2015}{\natexlab{b}})}\BibitemShut {NoStop}%
\bibitem [{\citenamefont {He}\ \emph {et~al.}(2011)\citenamefont {He},
  \citenamefont {Donadio},\ and\ \citenamefont {Galli}}]{he11}%
  \BibitemOpen
  \bibfield  {author} {\bibinfo {author} {\bibfnamefont {Yuping}\ \bibnamefont
  {He}}, \bibinfo {author} {\bibfnamefont {Davide}\ \bibnamefont {Donadio}}, \
  and\ \bibinfo {author} {\bibfnamefont {Giulia}\ \bibnamefont {Galli}},\
  }\bibfield  {title} {\enquote {\bibinfo {title} {Heat transport in amorphous
  silicon: Interplay between morphology and disorder},}\ }\href@noop {}
  {\bibfield  {journal} {\bibinfo  {journal} {Applied Physics Letters}\
  }\textbf {\bibinfo {volume} {98}},\ \bibinfo {eid} {144101} (\bibinfo {year}
  {2011})}\BibitemShut {NoStop}%
\bibitem [{\citenamefont {Larkin}\ and\ \citenamefont
  {McGaughey}(2014)}]{larkin14}%
  \BibitemOpen
  \bibfield  {author} {\bibinfo {author} {\bibfnamefont {Jason~M.}\
  \bibnamefont {Larkin}}\ and\ \bibinfo {author} {\bibfnamefont {Alan J.~H.}\
  \bibnamefont {McGaughey}},\ }\bibfield  {title} {\enquote {\bibinfo {title}
  {Thermal conductivity accumulation in amorphous silica and amorphous
  silicon},}\ }\href {\doibase 10.1103/PhysRevB.89.144303} {\bibfield
  {journal} {\bibinfo  {journal} {Phys. Rev. B}\ }\textbf {\bibinfo {volume}
  {89}},\ \bibinfo {pages} {144303} (\bibinfo {year} {2014})}\BibitemShut
  {NoStop}%
\bibitem [{\citenamefont {Lv}\ and\ \citenamefont {Henry}(2016)}]{lv15}%
  \BibitemOpen
  \bibfield  {author} {\bibinfo {author} {\bibfnamefont {Wei}\ \bibnamefont
  {Lv}}\ and\ \bibinfo {author} {\bibfnamefont {Asegun}\ \bibnamefont
  {Henry}},\ }\bibfield  {title} {\enquote {\bibinfo {title} {Direct
  calculation of modal contributions to thermal conductivity via
  {Green}--{Kubo} modal analysis},}\ }\href
  {http://stacks.iop.org/1367-2630/18/i=1/a=013028} {\bibfield  {journal}
  {\bibinfo  {journal} {New Journal of Physics}\ }\textbf {\bibinfo {volume}
  {18}},\ \bibinfo {pages} {013028} (\bibinfo {year} {2016})}\BibitemShut
  {NoStop}%
\bibitem [{\citenamefont {Zhou}\ and\ \citenamefont {Hu}(2015)}]{zhou15}%
  \BibitemOpen
  \bibfield  {author} {\bibinfo {author} {\bibfnamefont {Yanguang}\
  \bibnamefont {Zhou}}\ and\ \bibinfo {author} {\bibfnamefont {Ming}\
  \bibnamefont {Hu}},\ }\bibfield  {title} {\enquote {\bibinfo {title}
  {Quantitatively analyzing phonon spectral contribution of thermal
  conductivity based on nonequilibrium molecular dynamics simulations. {II.}
  {From} time {Fourier} transform},}\ }\href {\doibase
  10.1103/PhysRevB.92.195205} {\bibfield  {journal} {\bibinfo  {journal} {Phys.
  Rev. B}\ }\textbf {\bibinfo {volume} {92}},\ \bibinfo {pages} {195205}
  (\bibinfo {year} {2015})}\BibitemShut {NoStop}%
\bibitem [{\citenamefont {Turney}\ \emph {et~al.}(2009)\citenamefont {Turney},
  \citenamefont {McGaughey},\ and\ \citenamefont {Amon}}]{turney09a}%
  \BibitemOpen
  \bibfield  {author} {\bibinfo {author} {\bibfnamefont {J.~E.}\ \bibnamefont
  {Turney}}, \bibinfo {author} {\bibfnamefont {A.~J.~H.}\ \bibnamefont
  {McGaughey}}, \ and\ \bibinfo {author} {\bibfnamefont {C.~H.}\ \bibnamefont
  {Amon}},\ }\bibfield  {title} {\enquote {\bibinfo {title} {Assessing the
  applicability of quantum corrections to classical thermal conductivity
  predictions},}\ }\href {\doibase 10.1103/PhysRevB.79.224305} {\bibfield
  {journal} {\bibinfo  {journal} {Phys. Rev. B}\ }\textbf {\bibinfo {volume}
  {79}},\ \bibinfo {pages} {224305} (\bibinfo {year} {2009})}\BibitemShut
  {NoStop}%
\bibitem [{\citenamefont {Plimpton}(1995)}]{plimpton95}%
  \BibitemOpen
  \bibfield  {author} {\bibinfo {author} {\bibfnamefont {Steve}\ \bibnamefont
  {Plimpton}},\ }\bibfield  {title} {\enquote {\bibinfo {title} {Fast parallel
  algorithms for short-range molecular dynamics},}\ }\href {\doibase
  http://dx.doi.org/10.1006/jcph.1995.1039} {\bibfield  {journal} {\bibinfo
  {journal} {J. Comput. Phys.}\ }\textbf {\bibinfo {volume} {117}},\ \bibinfo
  {pages} {1 -- 19} (\bibinfo {year} {1995})}\BibitemShut {NoStop}%
\bibitem [{\citenamefont {Stillinger}\ and\ \citenamefont
  {Weber}(1985)}]{stillinger85}%
  \BibitemOpen
  \bibfield  {author} {\bibinfo {author} {\bibfnamefont {Frank~H.}\
  \bibnamefont {Stillinger}}\ and\ \bibinfo {author} {\bibfnamefont
  {Thomas~A.}\ \bibnamefont {Weber}},\ }\bibfield  {title} {\enquote {\bibinfo
  {title} {Computer simulation of local order in condensed phases of
  silicon},}\ }\href {\doibase 10.1103/PhysRevB.31.5262} {\bibfield  {journal}
  {\bibinfo  {journal} {Phys. Rev. B}\ }\textbf {\bibinfo {volume} {31}},\
  \bibinfo {pages} {5262--5271} (\bibinfo {year} {1985})}\BibitemShut {NoStop}%
\bibitem [{\citenamefont {Vink}\ \emph {et~al.}(2001)\citenamefont {Vink},
  \citenamefont {Barkema}, \citenamefont {van~der Weg},\ and\ \citenamefont
  {Mousseau}}]{vink01}%
  \BibitemOpen
  \bibfield  {author} {\bibinfo {author} {\bibfnamefont {R.L.C.}\ \bibnamefont
  {Vink}}, \bibinfo {author} {\bibfnamefont {G.T.}\ \bibnamefont {Barkema}},
  \bibinfo {author} {\bibfnamefont {W.F.}\ \bibnamefont {van~der Weg}}, \ and\
  \bibinfo {author} {\bibfnamefont {Normand}\ \bibnamefont {Mousseau}},\
  }\bibfield  {title} {\enquote {\bibinfo {title} {Fitting the
  {Stillinger}-{Weber} potential to amorphous silicon},}\ }\href {\doibase
  http://dx.doi.org/10.1016/S0022-3093(01)00342-8} {\bibfield  {journal}
  {\bibinfo  {journal} {Journal of Non-Crystalline Solids}\ }\textbf {\bibinfo
  {volume} {282}},\ \bibinfo {pages} {248--255} (\bibinfo {year}
  {2001})}\BibitemShut {NoStop}%
\bibitem [{\citenamefont {France-Lanord}\ \emph {et~al.}(2014)\citenamefont
  {France-Lanord}, \citenamefont {Blandre}, \citenamefont {Albaret},
  \citenamefont {Merabia}, \citenamefont {Lacroix},\ and\ \citenamefont
  {Termentzidis}}]{france-lanord14}%
  \BibitemOpen
  \bibfield  {author} {\bibinfo {author} {\bibfnamefont {Arthur}\ \bibnamefont
  {France-Lanord}}, \bibinfo {author} {\bibfnamefont {Etienne}\ \bibnamefont
  {Blandre}}, \bibinfo {author} {\bibfnamefont {Tristan}\ \bibnamefont
  {Albaret}}, \bibinfo {author} {\bibfnamefont {Samy}\ \bibnamefont {Merabia}},
  \bibinfo {author} {\bibfnamefont {David}\ \bibnamefont {Lacroix}}, \ and\
  \bibinfo {author} {\bibfnamefont {Konstantinos}\ \bibnamefont
  {Termentzidis}},\ }\bibfield  {title} {\enquote {\bibinfo {title} {Atomistic
  amorphous/crystalline interface modelling for superlattices and core/shell
  nanowires},}\ }\href {http://stacks.iop.org/0953-8984/26/i=5/a=055011}
  {\bibfield  {journal} {\bibinfo  {journal} {Journal of Physics: Condensed
  Matter}\ }\textbf {\bibinfo {volume} {26}},\ \bibinfo {pages} {055011}
  (\bibinfo {year} {2014})}\BibitemShut {NoStop}%
\bibitem [{\citenamefont {Chen}(2005)}]{chen}%
  \BibitemOpen
  \bibfield  {author} {\bibinfo {author} {\bibfnamefont {Gang}\ \bibnamefont
  {Chen}},\ }\href@noop {} {\emph {\bibinfo {title} {Nanoscale Energy Transport
  and Conversion}}}\ (\bibinfo  {publisher} {Oxford University Press, Oxford},\
  \bibinfo {year} {2005})\BibitemShut {NoStop}%
\bibitem [{\citenamefont {Rego}\ and\ \citenamefont
  {Kirczenow}(1998)}]{rego98}%
  \BibitemOpen
  \bibfield  {author} {\bibinfo {author} {\bibfnamefont {Luis G.~C.}\
  \bibnamefont {Rego}}\ and\ \bibinfo {author} {\bibfnamefont {George}\
  \bibnamefont {Kirczenow}},\ }\bibfield  {title} {\enquote {\bibinfo {title}
  {Quantized thermal conductance of dielectric quantum wires},}\ }\href
  {\doibase 10.1103/PhysRevLett.81.232} {\bibfield  {journal} {\bibinfo
  {journal} {Phys. Rev. Lett.}\ }\textbf {\bibinfo {volume} {81}},\ \bibinfo
  {pages} {232} (\bibinfo {year} {1998})}\BibitemShut {NoStop}%
\bibitem [{\citenamefont {Angelescu}\ \emph {et~al.}(1998)\citenamefont
  {Angelescu}, \citenamefont {Cross},\ and\ \citenamefont
  {Roukes}}]{angelescu98}%
  \BibitemOpen
  \bibfield  {author} {\bibinfo {author} {\bibfnamefont {D.~E.}\ \bibnamefont
  {Angelescu}}, \bibinfo {author} {\bibfnamefont {M.~C.}\ \bibnamefont
  {Cross}}, \ and\ \bibinfo {author} {\bibfnamefont {M.~L.}\ \bibnamefont
  {Roukes}},\ }\bibfield  {title} {\enquote {\bibinfo {title} {Heat transport
  in mesoscopic systems},}\ }\href {\doibase DOI: 10.1006/spmi.1997.0561}
  {\bibfield  {journal} {\bibinfo  {journal} {Superlattices Microstruct.}\
  }\textbf {\bibinfo {volume} {23}},\ \bibinfo {pages} {673 -- 689} (\bibinfo
  {year} {1998})}\BibitemShut {NoStop}%
\bibitem [{\citenamefont {S\"a\"askilahti}\ \emph {et~al.}(2013)\citenamefont
  {S\"a\"askilahti}, \citenamefont {Oksanen},\ and\ \citenamefont
  {Tulkki}}]{saaskilahti13}%
  \BibitemOpen
  \bibfield  {author} {\bibinfo {author} {\bibfnamefont {K.}~\bibnamefont
  {S\"a\"askilahti}}, \bibinfo {author} {\bibfnamefont {J.}~\bibnamefont
  {Oksanen}}, \ and\ \bibinfo {author} {\bibfnamefont {J.}~\bibnamefont
  {Tulkki}},\ }\bibfield  {title} {\enquote {\bibinfo {title} {Thermal balance
  and quantum heat transport in nanostructures thermalized by local {Langevin}
  heat baths},}\ }\href {\doibase 10.1103/PhysRevE.88.012128} {\bibfield
  {journal} {\bibinfo  {journal} {Phys. Rev. E}\ }\textbf {\bibinfo {volume}
  {88}},\ \bibinfo {pages} {012128} (\bibinfo {year} {2013})}\BibitemShut
  {NoStop}%
\bibitem [{\citenamefont {Mertig}\ \emph {et~al.}(1984)\citenamefont {Mertig},
  \citenamefont {Pompe},\ and\ \citenamefont {Hegenbarth}}]{mertig84}%
  \BibitemOpen
  \bibfield  {author} {\bibinfo {author} {\bibfnamefont {M.}~\bibnamefont
  {Mertig}}, \bibinfo {author} {\bibfnamefont {G.}~\bibnamefont {Pompe}}, \
  and\ \bibinfo {author} {\bibfnamefont {E.}~\bibnamefont {Hegenbarth}},\
  }\bibfield  {title} {\enquote {\bibinfo {title} {Specific heat of amorphous
  silicon at low temperatures},}\ }\href {\doibase
  http://dx.doi.org/10.1016/0038-1098(84)90589-1} {\bibfield  {journal}
  {\bibinfo  {journal} {Solid State Communications}\ }\textbf {\bibinfo
  {volume} {49}},\ \bibinfo {pages} {369 -- 372} (\bibinfo {year}
  {1984})}\BibitemShut {NoStop}%
\bibitem [{\citenamefont {Cahill}\ \emph {et~al.}(1994)\citenamefont {Cahill},
  \citenamefont {Katiyar},\ and\ \citenamefont {Abelson}}]{cahill94}%
  \BibitemOpen
  \bibfield  {author} {\bibinfo {author} {\bibfnamefont {David~G.}\
  \bibnamefont {Cahill}}, \bibinfo {author} {\bibfnamefont {M.}~\bibnamefont
  {Katiyar}}, \ and\ \bibinfo {author} {\bibfnamefont {J.~R.}\ \bibnamefont
  {Abelson}},\ }\bibfield  {title} {\enquote {\bibinfo {title} {Thermal
  conductivity of \textit{a}-{Si}:{H} thin films},}\ }\href {\doibase
  10.1103/PhysRevB.50.6077} {\bibfield  {journal} {\bibinfo  {journal} {Phys.
  Rev. B}\ }\textbf {\bibinfo {volume} {50}},\ \bibinfo {pages} {6077--6081}
  (\bibinfo {year} {1994})}\BibitemShut {NoStop}%
\end{thebibliography}
\end{document}